\documentclass[a4paper,english]{IEEEtran}
\usepackage{graphicx}
\usepackage{amsmath}
\usepackage{amssymb}
\usepackage{algorithmic}
\usepackage{algorithm}
\usepackage[dvipdfm]{hyperref}

\usepackage{pifont}
\usepackage{graphics,graphicx,psfrag,color,pstcol,pst-grad}

\definecolor{hhicyan}{cmyk}{0.19,0,0,0}
\definecolor{hhiyellow}{cmyk}{0,0,0.20,0}
\definecolor{myyellow}{cmyk}{0,0,.7,0}

\newcommand{\Trace}[1]{\mathbf{Tr}{#1}}
\newcommand{\setrankone}{Z}
\newcommand{\settraceideal}{M_1}
\newcommand{\settraceclass}{\mathcal{T}_1}

\newcommand{\order}{o}

\newcommand{\defeq}{\overset{\text{def}}{=}}

\newcommand{\Bx}{\boldsymbol{x}}

\newcommand{\Bn}{\boldsymbol{n}}
\newcommand{\HH}{{\mathcal{H}}}
\newcommand{\BH}{\boldsymbol{\HH}}

\newcommand{\sigmaN}{{\sigma^2}}

\newcommand{\BHSpread}{\boldsymbol{\Sigma}}

\newcommand{\BHScat}{\boldsymbol{C}}

\newcommand{\major}{\succ}

\newcommand{\Ltwo}{{\mathcal{L}}_2}
\newcommand{\Lone}{L_1}

\newcommand{\Indexset}{{\mathcal{I}}}

\newcommand{\Id}{{\mathbb{I}}}

\newcommand{\Amb}{{\mathbf{A}}}

\newcommand{\SINR}{\text{\small{\rm{SINR}}}}

\newcommand{\EX}[1]{{\mathbf{E}}\{#1\}}
\newcommand{\Ex}[2]{{\mathbf{E}}_{#1}\{#2\}}

\newcommand{\Shift}{{\boldsymbol{S}}}

\newcommand{\Position}{{\boldsymbol{X}}}
\newcommand{\Momentum}{{\boldsymbol{D}}}
\newcommand{\Schwarz}{{\mathcal{S}}}

\newcommand{\fourier}[1]{\hat{#1}}

\newcommand{\opt}{\text{\rm (opt)}}

\begin{document}
\title{A Group-Theoretic Approach to the WSSUS Pulse Design Problem}
\author{Peter Jung and Gerhard Wunder \\
  Sino-German Mobile Communications Institute (MCI) \\[.1em]
  \small{\{jung,wunder\}@hhi.fraunhofer.de}}
\maketitle
\begin{abstract}
   We consider the pulse design problem in multicarrier transmission where the pulse shapes 
   are adapted to the second order statistics of the WSSUS channel.
   Even though the problem has been addressed by many authors analytical insights 
   are rather limited. 
   First we show that the problem is equivalent to the pure state channel fidelity in quantum
   information theory.
   Next we present a new approach where the original optimization functional is 
   related to an eigenvalue
   problem for a pseudo differential operator by utilizing unitary representations of the Weyl--Heisenberg group.  
   A local approximation of the operator for underspread channels is derived which implicitly covers the 
   concepts of pulse scaling and optimal phase space displacement. The problem is reformulated as 
   a differential equation and the optimal pulses occur as eigenstates of the harmonic oscillator 
   Hamiltonian. Furthermore this
   operator--algebraic approach is extended to provide exact solutions for different classes of
   scattering environments.
\end{abstract}

\section{Introduction}
Pulse shaping in multicarrier transmission is a key ingredient for high rate wireless links. 
Furthermore it is the standard tool to mitigate the interference caused by doubly dispersive channels.
Most multicarrier schemes like conventional OFDM exploiting guard regions (a cyclic prefix), 
pulse shaped OFDM and OFDM/OQAM can be jointly formulated.
Hence we focus on a transmit baseband signal $s(t)$ given as
\begin{equation}
   \begin{aligned}
      s(t)
      &=\sum_{(mn)\in\Indexset}x_{mn}e^{i2\pi mFt}\gamma(t-nT)
      =\sum_{(mn)\in\Indexset}x_{mn}\gamma_{mn}(t)
   \end{aligned}
   \label{equ:txsignal}
\end{equation}
where $i$ is the imaginary unit and 
$\gamma_{mn}\defeq\Shift_{(nT,mF)}\,\gamma$ are time-frequency shifted 
versions of the transmit pulse $\gamma$, i.e. shifted according to the lattice
$T\mathbb{Z}\times F\mathbb{Z}$. It is also beneficial to consider different lattice
structures \cite{strohmer:lofdm2} on which our contribution will apply as well.
The time-frequency (or phase space) shift operator $\Shift_{(\tau,\nu)}$ is intimately
connected to unitary representations of the Weyl-Heisenberg group as we will elaborate later on.
Therefore (\ref{equ:txsignal}) is also known as Weyl-Heisenberg or Gabor signaling. 

The coefficients $x_{mn}$ in (\ref{equ:txsignal}) are the complex 
data symbols at time instant $n$ and subcarrier index $m$ with the property
$\EX{\Bx\Bx^*}=\Id$ ($\cdot^*$ means conjugate transpose) where \mbox{$\Bx=(\dots,x_{mn},\dots)^T$}.
The indices $(mn)$ range over some doubly-countable index set $\Indexset$, 
referring to the data burst to be transmitted. 
We will denote the linear time-variant channel by $\BH$  and the 
additive white Gaussian noise process (AWGN) by $n(t)$.
The received signal is then
\begin{equation}
   \begin{aligned}
      r(t)=(\BH s)(t)+n(t)
      =\iint \BHSpread(\tau,\nu)(\Shift_{(\tau,\nu)}s)(t)d\tau d\nu + n(t)
   \end{aligned}
   \label{eq:rxsignal}
\end{equation}
with $\BHSpread(\tau,\nu)$ being a realization of the ''channel spreading
function''. In practice  $\BHSpread(\tau,\nu)$ is causal and has finite support. 
We used here the notion of
the WSSUS channel. In the WSSUS assumption the
channel is characterized by the second order statistics of $\BHSpread(\tau,\nu)$, i.e.
\begin{equation*}
   \EX{\BHSpread(\tau,\nu)
     \overline{\BHSpread(\tau',\nu')}}=\BHScat(\tau,\nu)\delta(\tau-\tau')\delta(\nu-\nu')
\end{equation*}
where $\BHScat(\tau,\nu)$ is the scattering function.
Without loss of generality we assume $\lVert\BHScat\rVert_1=1$.
To obtain the data symbol $\tilde{x}_{kl}$ the receiver does the
projection on $g_{kl}\defeq\Shift_{(lT,kF)}g$, i.e.
\begin{equation*}
   \tilde{x}_{kl}=\langle g_{kl},r\rangle=\int\overline{g}_{kl}(t)r(t)dt
\end{equation*} 
By introducing the elements $H_{kl,mn}\defeq\langle g_{kl},\BH\gamma_{mn}\rangle$ of the
channel matrix $H\in\mathbb{C}^{\Indexset\times\Indexset}$, 
the multicarrier transmission can be formulated as the linear equation
\mbox{$\tilde{\Bx}=H\Bx+\tilde{\Bn}$},
where $\tilde{\Bn}$ is the vector of the projected noise having a
power of $\sigmaN$ per component. 
We assume that the receiver has perfect channel knowledge
(given by $\BHSpread(\tau,\nu)$), i.e.
single carrier based equalization in the absence of noise would be
$\tilde{x}^{\text{eq}}_{kl}=\tilde{x}_{kl}/H_{kl,kl}$, with
\begin{equation*}
   \begin{aligned}
      H_{kl,kl}
      &=\langle g_{kl},\BH\gamma_{kl}\rangle 
      =\iint \BHSpread(\tau,\nu)\langle g_{kl},\Shift_{(\tau,\nu)}\gamma_{kl}\rangle d\tau d\nu\\
      &\defeq\iint \BHSpread(\tau,\nu)e^{-i2\pi(\tau kF-\nu lT)}\Amb_{g\gamma}(\tau,\nu) d\tau d\nu
   \end{aligned}
\end{equation*}
where $\Amb_{g\gamma}(\tau,\nu)=\langle g,\Shift_{(\tau,\nu)}\gamma\rangle$ 
is the cross ambiguity function of the pulse pair $\{g,\gamma\}$.

\section{Problem Statement}
Considering only single carrier equalization, it is
natural to require $a\defeq|H_{kl,kl}|^2$ (the channel gain) to be maximal and the interference power 
$b\defeq\sum_{(kl)\neq(mn)}|H_{kl,mn}|^2$ to be minimal as possible. This addresses the concept of 
{\it pulse shaping}. However to be practicable, 
the pulses should be adapted to the second order statistics only, given by $\BHScat(\tau,\nu)$ 
and {\bf not} to a particular channel realization $\BHSpread(\tau,\nu)$. 
Hence, we aim at maximization of 
\begin{equation*}
   \SINR\defeq{\frac{\Ex{\BH}{a}}{\sigmaN+\Ex{\BH}{b}}}
\end{equation*}
by proper design of $\gamma$ and $g$. Up to very few 
special cases the analytical solution of this global optimization problem (jointly non-convex in $(\gamma,g)$)
is unknown. However numerical optimization methods 
are presented in \cite{schafhuber:pimrc02,schniter:allerton03,jung:spawc2004}. Following our previous work \cite{jung:spawc2004} we simplify 
the problem by proposing a relaxation, which separates the problem into two steps.
Upper bounding \mbox{$\Ex{\BH}{b}\leq B_\gamma - \Ex{\BH}{a}$} gives a lower bound on
$\SINR$ (see \cite{jung:spawc2004}), where $B_\gamma$ is the so called Bessel bound of $\{\gamma_{mn}\}$
\cite{christensen:framesandrieszbases}. 
In this paper we focus on the first step only where $\Ex{\BH}{a}$ should be maximized. This gives
the following optimization problem
\begin{equation}
   \begin{aligned}
      \{\gamma^{\opt},g^{\opt}\}
      &=\arg\max{\Ex{\BH}{a}}\\
      &=\arg\max_{\lVert\gamma\rVert_2=\lVert g\rVert_2=1}{\int|\Amb_{g\gamma}(\tau,\nu)|^2d\mu}
    \end{aligned}      
    \label{eq:gainoptimization}
\end{equation}
where $d\mu\defeq\BHScat(\tau,\nu)d\tau d\nu$.
In this context it was first introduced in \cite{kozek:nofdm1} respectively 
\cite{kozek:thesis}, but similar problems already
occurred in radar literature much earlier. 
In particular for the elliptical symmetry of $\BHScat(\tau,\nu)$ Hermite functions
establish local extremal points as found in \cite{kozek:thesis}. The
scaling rule for fixed pulses was studied in \cite{kozek:thesis,liu:orthogonalstf}.
Also it is possible to find a 
close relation to the channel fidelity and minimum output entropy states 
in quantum information theory as we will show later on.
In particular the important class of Gaussian scattering profiles (corresponding to classical 
bosonic quantum channels) was already addressed in \cite{arxiv:0404005} and
\cite{arxiv:0409063}. 

Out of the scope of this paper is the second step, in which the minimization of $B_\gamma$
(which depends on $\gamma^{\opt}$) is achieved. This well known procedure \cite{strohmer:lofdm2} 
(in the case of $TF>1$ that is to find the ''nearest'' orthogonal Gabor basis with respect to the $\Ltwo$-norm) is also described in \cite{jung:spawc2004}.
Unfortunately the resulting pulses will be in general again a suboptimal solution of ($\ref{eq:gainoptimization}$).
See \cite{strohmer:dualgaborframes} for a discussion of this problem.
Nevertheless, this separation and therefore (\ref{eq:gainoptimization}) opens up
analytical insights into the pulse design problem.

\section{Contributions}
The original formulation of the pulse design problem in (\ref{eq:gainoptimization}) 
hides the internal group structure induced by the time-frequency shift operators.
In this paper we derive a lower bound for the optimization
functional (\ref{eq:gainoptimization}) on which we can exploit this structure explicitely.
Moreover we sketch that the results will hold in the direct problem with minor restrictions.
We present an operator--algebraic reformulation by utilizing representation theory 
of the Weyl--Heisenberg group. Our approach relates the optimal pulses to approximate 
eigenstates of pseudo differential operators. The procedure naturally embeds
the concepts of pulse scaling and optimal time-frequency offsets (or phase space displacement).
Then we extent our framework to
provide exact solutions for the class of Gaussian scattering profiles.
Because the underlying theory is partially  
not very common in multicarrier community we will give a short introduction to the few properties
we will need for our investigation. More details can be found in \cite{folland:harmonics:phasespace}.

\subsection{The Weyl-Heisenberg Group and Pseudo differential Operators}
The two families of shift operators $\Shift_{(\tau,0)}$ and $\Shift_{(0,\nu)}$
are unitary representations of the group corresponding to 
the real line $\mathbb{R}$  with addition as group operation. The extension to $\mathbb{R}^2$ in the sense 
of 
\begin{equation}
   \Shift_{(\alpha,\beta)}\cdot \Shift_{(\gamma,\delta)}=e^{-i2\pi\alpha\delta}\Shift_{(\alpha+\gamma,\beta+\delta)}
   \label{eq:shift:commutationrule}
\end{equation}
is not closed because of the phase factor.
Closeness is achieved by introducing the torus ($\mathbb{T}$) as the  third variable, i.e.
\begin{equation*}
   e^{i2\pi\phi}\Shift_{(\alpha,\beta)}\cdot e^{i2\pi\psi}\Shift_{(\gamma,\delta)}=
   e^{i2\pi(\phi+\psi-\alpha\delta)}\Shift_{(\alpha+\gamma,\beta+\delta)}
\end{equation*}
The corresponding group $\mathbb{H}=\mathbb{R}\times\mathbb{R}\times\mathbb{T}$ with the group law
$(\alpha,\beta,\phi)(\gamma,\delta,\psi)=(\alpha+\gamma,\beta+\delta,\phi+\psi-\alpha\delta)$ is called
the (reduced\footnote{The addition in third component is taken to be 
  mod $1$. Otherwise this yields the (full) polarized Heisenberg group with non-compact center.}) 
polarized {\it Heisenberg group} (HG). 
The HG can be represented as a group of upper triangular matrices by the 
group homomorphism
\begin{equation*}
   (\alpha,\beta,\phi)\rightarrow   
   H(\alpha,\beta,\phi)=
   \left(\begin{array}{ccc}
        1 & \alpha &  \phi \\
        0 &      1 & \beta \\
        0 &      0 &     1 \\
     \end{array}\right)
\end{equation*}
where the group action is matrix multiplication.
The matrices $h(\alpha,\beta,\phi)=H(\alpha,\beta,\phi)-1$ written with 
$d=(1,0,0)$, $x=(0,1,0)$ and $e=(0,0,1)$
as $h(\alpha,\beta,\phi)=\alpha h(d)+\beta h(x)+\phi h(e)$ are clearly isomorphic to 
$\mathbb{R}^3$ and with the matrix commutator they turn into a Lie algebra. The Lie bracket 
in this case is $\left[(\alpha,\beta,\phi),(\gamma,\delta,\psi)\right]\defeq(0,0,\alpha\delta-\beta\gamma)$.
Due to the bilinearity of the Lie bracket this can be shortly written as the
{\it Heisenberg Commutation Relations}, i.e.
\mbox{$\left[d,x\right]=e\,\,\,\left[x,e\right]=0\,\,\,\left[d,e\right]=0$}.
That this is exactly the Heisenberg algebra connected to the HG follows from
$h(\alpha,\beta,\phi)^2=h(0,0,\alpha\beta)$ and $h(\alpha,\beta,\phi)^n=0$ for $n>2$. 
The exponential map of the matrix $h(\alpha,\beta,\phi)$ is then given as
\begin{equation}
   \begin{split}
      e^{h(\alpha,\beta,\phi)}
      &=\sum_{n=0}^\infty\frac{h(\alpha,\beta,\phi)^n}{n!}
      =1+h(\alpha,\beta,\phi)+\frac{1}{2}h(0,0,\alpha\beta)\\
      &= H(\alpha,\beta,\phi+\frac{1}{2}\alpha\beta)
      \raisetag{2em}
   \end{split}
   \label{eq:heisenberg:nilpotent}
\end{equation}
Thus, it maps the Heisenberg algebra to the unpolarized HG. 
The series expansion is finite (the elements $h(\alpha,\beta,\phi)$ are nilpotent 
endomorphisms). 
Returning to the polarized Heisenberg group we transform finally
\mbox{$H(\alpha,\beta,\phi)=H(0,0,-\frac{1}{2}\alpha\beta)e^{h(\alpha,\beta,\phi)}$}.

To establish the connection to $\Shift_{(\alpha,\beta)}$ considered as operators on $\Schwarz(\mathbb{R})$
(the Schwartz space of rapidly decreasing functions) we have to switch to the so called
{\it Schr\"odinger representation}.
In this picture the hermitian operators $\Position$ and $\Momentum$ with
\begin{equation*}
   \begin{split}
      (\Position f)(t)&\defeq tf(t) \\
      (\Momentum f)(t)&\defeq \frac{1}{2\pi i}f'(t)
   \end{split}
\end{equation*}
setup a basis representation for the Heisenberg Lie algebra. 
The skew-hermitian operators $2\pi i\Position$ 
(generates the frequency shifts), $2\pi i\Momentum$ (generates the time shifts) 
and $2\pi i\boldsymbol{E}=2\pi i$ ($E$ is the identity) correspond 
to $x,d$ and $e$. They give again the Heisenberg commutation rules, 
hence linear combinations of them fulfill the same Lie bracket (the commutator of linear operators)  and consequently 
\mbox{$(\tau,\nu,s)\rightarrow d\rho(\tau,\nu,s)=2\pi i(s+\nu\Position+\tau\Momentum)$}
is again a Lie algebra isomorphism for the Heisenberg algebra.
As in (\ref{eq:heisenberg:nilpotent}) the HG is then given by exponentiation, i.e. the so called Weyl operator is 
given as
\begin{equation*}
   \begin{aligned}
      \rho(\tau,\nu,s)
      &=e^{d\rho(\tau,\nu,s)}=e^{2\pi i(s+\nu\Position+\tau\Momentum)}
      =e^{2\pi is}e^{\pi i\tau\nu}\Shift_{(-\tau,\nu)}
   \end{aligned}
\end{equation*}
With $\rho(\tau,\nu)\defeq\rho(\tau,\nu,0)$ we have
$\Shift_{(\tau,\nu)}=e^{\pi i\tau\nu}\rho(-\tau,\nu)=e^{\pi i\tau\nu}e^{2\pi i(\nu\Position-\tau\Momentum)}$, i.e.
integrals over shift operators as in (\ref{eq:rxsignal}) are in fact pseudo differential operators \cite{folland:harmonics:phasespace}
of the following spreading representation (the Weyl transform)
\begin{equation*}
   \begin{aligned}
      \sigma(\Momentum,\Position)=\iint \fourier{\sigma}(\tau,\nu)e^{2\pi i(\nu\Position+\tau\Momentum)}d\tau d\nu
   \end{aligned}
\end{equation*}
$\fourier{\sigma}(\tau,\nu)$ is called the spreading function (or representing function, i.e. 
the 2D Fourier transform of the symbol $\sigma(d,x)$ of the operator $\sigma(\Momentum,\Position)$). 

\subsection{The WSSUS Pulse Design Problem}
Straight forward calculation
shows now that the squared magnitude of the cross ambiguity function
$|\Amb_{g\gamma}(\tau,\nu)|^2$ can be written in the following form
\begin{equation}
   \begin{aligned}
      |\Amb_{g\gamma}(\tau,\nu)|^2
      &=\langle g,\Shift_{(\tau,\nu)}\gamma\rangle\langle\gamma,\Shift^*_{(\tau,\nu)}g\rangle\\
      &=\Trace{\,G\Shift_{(\tau,\nu)}\Gamma\Shift^*_{(\tau,\nu)}}
   \end{aligned}
\end{equation}
where $G$ ($\Gamma$) is the (rank-one) orthogonal projector onto $g$ ($\gamma$).
By that transformation we emphasize that $\Gamma$ undergoes a linear transformations
before being projected onto $g$. This special kind of linear transformation is also
called a unitary evolution, which preserve the spectrum of $\Gamma$ (in our
case the rank). This obviously does not hold in generality if taking the sum over
different unitary evolution of the same argument. 
Hence, we collect them together by defining affine maps $A$ and $\tilde{A}$ such that
\begin{equation}
   \begin{aligned}
      \Ex{\BH}{a}
      &=\Trace{\,G[\int\Shift_{(\tau,\nu)}\Gamma\Shift^*_{(\tau,\nu)}}d\mu]
      \defeq\Trace{\,G A(\Gamma)}\\
      &=\Trace{\,\Gamma[\int\Shift^*_{(\tau,\nu)}G\Shift_{(\tau,\nu)}}d\mu]
      \defeq\Trace{\,\Gamma\tilde{A}(G)}\\
   \end{aligned}
   \label{eq:cpmap:introduction}
\end{equation}
The main reason for this reformulation is the notion of {\it completely positive
  maps} (CP-maps) \cite{stinespring:positivefunctions} 
which directly apply on the pulse design problem.
CP-maps  like $A(\cdot)$
received much attention due to its application in quantum 
information theory.
Before going more in detail, let us define $\settraceclass$ as the set
of trace class operators. The set
\mbox{$\settraceideal\defeq\{z\,|\,z\in\settraceclass,z=z^*,z\geq0,\Trace{\,z}=1\}$}
is a convex subset of $\settraceclass$. With $\setrankone$ we will denote the extremal 
boundary of $\settraceideal$, which is  the set of all orthogonal rank-one projectors. 
With the definition of $\tilde{A}$ in (\ref{eq:cpmap:introduction}) follows that
$\tilde{A}$ is adjoint of $A$ with respect to the inner product $\Trace{X^*Y}$.
Due to $\lVert\BHScat\rVert_1=1$ both maps are trace preserving $\Trace{A(X)}=\Trace{X}$.
Moreover they are hermiticity preserving $A(X)^*=A(X^*)$ and 
entropy increasing $X\major A(X)$ ($\major$ is the partial
order due to eigenvalue majorization). 
The complete positivity and the trace-preserving property is ensured by 
\begin{equation*}
   \begin{aligned}
      \int d\mu \Shift^*_{(\tau,\nu)}\Shift_{(\tau,\nu)}=\Id
   \end{aligned}
\end{equation*}
With this framework we can write now the optimization problem as
\begin{equation}
   \begin{aligned}
      \max_{G,\Gamma\in\setrankone}{\Trace{\,G A(\Gamma)}}
   \end{aligned}
   \label{eq:optimization:trace}
\end{equation}
where $\Gamma$ represent the transmitter and the CP-map $A(\cdot)$ represent
the ''averaged'' action of the channel and $G$ is the receiver.
This formulation is similar to the channel fidelity in quantum information
processing. In fact - the problems are equivalent if considering so called 
pure states. The initial preparation of a pure quantum state (the symbol to transmit)
is represented by a so called rank-one density operator (in our case $\Gamma$). The quantum
channel is represented by a CP-map $A(\cdot)$ having again a density operator as its output.
The measurement (the detection of the transmitted symbol) is performed in our case with $G$.
Obviously either $G$ or $\Gamma$ can be dropped in the optimization, i.e.
\begin{equation}
   \begin{aligned}
      \max_{G,\Gamma\in\setrankone}{\Trace{\,G A(\Gamma)}}
      =\max_{\Gamma\in\setrankone}{\lVert A(\Gamma)\rVert_\infty}
      =\max_{G\in\setrankone}{\lVert \tilde{A}(G)\rVert_\infty}
   \end{aligned}
   \label{eq:reformulation:pulsedesign:rankone}
\end{equation}
where $\lVert\cdot\rVert_\infty$ denotes the operator norm.
This measure represents the maximum achievable 
purity of the output of a quantum channel with pure states as input.
Turning back to the language of WSSUS signaling, this represents the maximum achievable 
''energy'' which can be collected by a single pulse if communicating with the optimal 
pulse $\gamma$ over a large ensemble of WSSUS channels. CP-maps over the Heisenberg group 
have some more important properties. One is the covariance property with respect to
group elements which follows from (\ref{eq:shift:commutationrule}), i.e.
\begin{equation}
   A(\Shift_{(\tau,\nu)}\Gamma\Shift^*_{\tau,\nu})=\Shift_{(\tau,\nu)}A(\Gamma)\Shift^*_{(\tau,\nu)}
   \label{eq:covariance}
\end{equation}
The physical meaning is that (\ref{eq:optimization:trace}) is invariant with respect to 
common time--frequency shifts of $G$ and $\Gamma$. A trivial but important conclusion is that Weyl-Heisenberg (Gabor)
signaling is a reasonable scheme, which guarantees the same performance on all lattice points.
Alternatively it can be viewed in the quantum picture as symbol alphabet
of pure states achieving all the same fidelity.
From (\ref{eq:covariance}) follows furthermore that different maps $A_1$ and $A_2$ commute, i.e.
$A_1\circ A_2=A_2\circ A_1$.
Coming back to the formulation of pulse design problem in (\ref{eq:reformulation:pulsedesign:rankone}) 
we can finally relax the constraint set from $\setrankone$ to $\settraceideal$ which gives 
\begin{equation}
   \begin{aligned}
      \max_{\Gamma\in\settraceideal}{\lVert A(\Gamma)\rVert_\infty}
      =\max_{G\in\settraceideal}{\lVert \tilde{A}(G)\rVert_\infty}
   \end{aligned}
\end{equation}
provided by the convexity of $\lVert\cdot\rVert_\infty$ and linearity of $A(\cdot)$.
To the authors knowledge this reformulation of the pulse design criterion 
as a convex maximization problem seems to be new.
Without further investigations of the analytical structure of $A(\cdot)$ 
such global-type optimization problems are in general difficult to solve.
Therefore we will emphasize in the following more on the Heisenberg group structure contained in 
$A(\cdot)$.

\subsection{The Schr\"odinger Representation}
The connection between Weyl operators (the unitary representations of the Weyl-Heisenberg
group in the Schr\"odinger picture) and $\Shift_{(\tau,\nu)}$ will reveal the fundamental role
of Gaussians in WSSUS signaling. We will show this first in a simpler lower bound analysis 
which mainly admits the same maximizer as the original problem (given in the appendix).
Thus, coming back now to (\ref{eq:gainoptimization}) and let  
$(\tau_0,\nu_0)$ be an arbitrary offset between $g$ and $\gamma$ in the time-frequency plane,
hence we define $\tilde{\gamma}=\Shift_{(\tau_0,\nu_0)}\gamma$. 
\begin{equation}
   \begin{split}
      \Ex{\BH}{a}
      &=\int |\Amb_{g\gamma}(\tau,\nu)|^2d\mu
      =\int |\langle g,\Shift_{(\tau-\tau_0,\nu-\nu_0)}\tilde{\gamma}\rangle|^2d\mu\\
      &=\int |\langle g,\rho(-\tau+\tau_0,\nu-\nu_0)\tilde{\gamma}\rangle|^2d\mu\\
      &\geq(\int|\langle g,\rho(-\tau+\tau_0,\nu-\nu_0)\tilde{\gamma}\rangle| d\mu)^2\\
      &\geq|\int\langle g,\rho(-\tau+\tau_0,\nu-\nu_0)\tilde{\gamma}\rangle d\mu|^2\\
      &=|\langle g,[\int\rho(-\tau+\tau_0,\nu-\nu_0)d\mu]\tilde{\gamma}\rangle |^2
      \defeq|\langle g,\mathcal{L}\tilde{\gamma}\rangle |^2
      \raisetag{6em}
   \end{split}
   \label{eq:lowerbound}
\end{equation}
In the latter we used Jensen's inequality\footnote{It can be shown that $\Amb_{g\gamma}\in\Lone(\mu)$.}
($\int d\mu=\lVert\BHScat\rVert_1=1$, see also \cite{jung:spawc2004}).
We will use now (\ref{eq:lowerbound}) for further analytical studies. The bound 
becomes sharp iff $\xi\Amb_{g\gamma}(\tau,\nu)\in\mathbb{R}$ is constant on $\text{supp}\,\BHScat$
for some $\xi\in\mathbb{T}$, hence is well suited for
underspread channels.
The operator $\mathcal{L}$ is a pseudo differential operator with spreading function 
$\fourier{\sigma}(\tau,\nu)=-\BHScat(\tau_0-\tau,\nu+\nu_0)$. 
\begin{equation*}
   \mathcal{L}=\iint-\BHScat(\tau_0-\tau,\nu+\nu_0)e^{2\pi i(\nu\Position+\tau\Momentum)}d\tau d\nu
\end{equation*}

\noindent{\bf Local approximation:}
The nilpotent property with respect to the matrix product celebrated in (\ref{eq:heisenberg:nilpotent}) 
unfortunately does not translate into the Schr\"odinger picture, so that 
\begin{equation}
   \begin{split}
      \Shift_{(\tau,\nu)}
      &=e^{\pi i\tau\nu}\rho(-\tau,\nu)\\
      &=e^{\pi i\tau\nu}[1+d\rho(-\tau,\nu)+\frac{1}{2}d\rho(-\tau,\nu)^2]+\order(2)\\
      &\approx e^{\pi i\tau\nu}[1+2\pi iK-2\pi^2K^2]
      \raisetag{3.5em}
      \vspace*{1em}
   \end{split}
   \label{eq:shift:localapprox}
\end{equation}
with the hermitian operator $K\defeq\nu\Position-\tau\Momentum$ holds only as an approximation 
(for $\tau$ and $\nu$ being small), i.e. gives a local approximation of $\mathcal{L}$ which is
\begin{equation*}
   \begin{aligned}
      L\defeq C_{00}
      &+2\pi i(C_{01}\Position-C_{10}\Momentum)\\
      &-2\pi^2(C_{02}\Position^2+C_{20}\Momentum^2-C_{11}[\Position\Momentum+\Momentum\Position])\\
   \end{aligned}      
\end{equation*}
where $C_{mn}=\iint\BHScat(\tau,\nu)(\tau-\tau_0)^m(\nu-\nu_0)^n$ 
are the moments of the scattering function around $(\tau_0,\nu_0)$.
Because $\Position$ and $\Momentum$ are hermitian operators, $L$ is hermitian too if 
$C_{mn} i^{m+n}\in\mathbb{R}$ for
$m,n=0,1,2$. In this case the optimization problem is an eigenvalue problem.
Moreover then it follows that $g=\alpha L\tilde{\gamma}$ for some $\alpha\in\mathbb{C}$, because only in this case
equality in $|\langle g,L\tilde{\gamma}\rangle|\leq\lVert g\rVert_2\lVert L\tilde{\gamma}\rVert_2$
is achieved. $L$ can be made hermitian if we choose 
$\tau_0=\lVert\tau\BHScat\rVert_1$  and $\nu_0  =\lVert \nu\BHScat\rVert_1$, so
that $C_{10}=C_{01}=0$. Thus we have
\begin{equation*}
   L= C_{00}-2\pi^2[C_{02}\Position^2+C_{20}\Momentum^2-C_{11}(\Position\Momentum+\Momentum\Position)]
\end{equation*}
which is an hermitian differential operator of second order. 
With 
\begin{equation*}
   (d_\alpha f)(t)\defeq \frac{1}{\sqrt{\alpha}}f(t/\alpha)
\end{equation*}
we define now dilated functions 
\mbox{$g_\alpha\defeq d_\alpha g$}, \mbox {$\tilde{\gamma}_\alpha\defeq d_\alpha\tilde{\gamma}$}
and the dilated operator $L_\alpha\defeq d_\alpha L d_{1/\alpha}$.
Using furthermore that 
\begin{equation}
   \begin{split}
      d_\alpha\Position d_{1/\alpha}&=\frac{1}{\alpha}\Position\\
      d_\alpha\Momentum d_{1/\alpha}&=\alpha\Momentum\\
   \end{split}
   \label{eq:operator:dilation}
\end{equation}
we get
\begin{equation*}
   \begin{aligned}
      \langle g_\alpha,L_\alpha\tilde{\gamma}_\alpha\rangle
      =\langle g_\alpha, C_{00}&-2\pi^2[\frac{C_{02}}{\alpha^2}\Position^2+C_{20}\alpha^2\Momentum^2\\
      &-C_{11}(\Position\Momentum+\Momentum\Position)]
      \tilde{\gamma}_\alpha\rangle 
   \end{aligned}
\end{equation*}
and with $\alpha^4=C_{02}/C_{20}$ the phase space symmetric version
\begin{equation*}
   \begin{aligned}
      L_\alpha=2\pi^2\sqrt{C_{02}C_{20}}\{\kappa-[\Position^2 + \Momentum^2-
      C_{11}(\Position\Momentum+\Momentum\Position)]\}
   \end{aligned}
\end{equation*}
where the constant is $\kappa=\frac{C_{00}}{2\pi^2\sqrt{C_{02}C_{20}}}$ (with 
our WSSUS assumptions follows also $C_{00}=1$). 
For simplicity let us assume that the shifted scattering
function is separable yielding $C_{11}=0$. In the general case the $C_{11}$-term 
can be removed using a proper symplectic transformation 
(see for example \cite{taylor:noncommutative:harmonic:analysis}).
The eigenfunction of the so called sub-Laplacian (or the harmonic oscillator 
Hamiltonian) $\Position^2 + \Momentum^2$ 
are the Hermite functions $h_n$, with $(\Position^2 + \Momentum^2)h_n=\frac{2n+1}{2\pi}h_n$.
Therefore it follows that 
\begin{equation*}
   L_\alpha h_n=(C_{00}-\pi\sqrt{C_{02}C_{20}}(2n+1))h_n
\end{equation*}
Hence in local approximation the maximization problem is solved by
$h_0$, i.e. $g=d_{1/\alpha}h_0$ and $\gamma=\Shift^{-1}_{(\tau_0,\nu_0)}d_{1/\alpha}h_0$
which are both  scaled and proper separated Gaussians (the ground state of the harmonic oscillator).
This is an important (and expected) result for the pulse design problem in WSSUS channels. It includes
the concepts of pulse scaling (by $d_{1/\alpha}$) and proper phase space displacement 
(by $\Shift^{-1}_{(\tau_0,\nu_0)}$) as natural operations.
However this approximations is only valid for \mbox{$C_{02}C_{20}\ll 1$} (underspread channel), such that
$(C_{00}-\pi\sqrt{C_{02}C_{20}}(2n+1))>0$. We obtain the same solutions in the original problem 
if we apply this approximation (see the appendix). Next we will derive cases where this approximation turns out to be
is exact. \\[1em]
\noindent{\bf Gaussian scattering functions:}
Let us assume that after performing proper pulse scaling and separation the scattering function is given 
as the symmetric Gaussian $\BHScat(\tau,\nu)=\frac{\alpha}{2}e^{-\frac{\pi}{2}\alpha(\tau^2+\nu^2)}$
where $0<\alpha\in\mathbb{R}$. If $\alpha\gg 1$ the channel is underspread.
It can be shown that then $\mathcal{L}$ essentially self--adjoint, hence the maximum in
(\ref{eq:lowerbound}) is again achieved by eigenfunctions of $\mathcal{L}$. Operators having such spreading functions
are contained in the so called {\it oscillator semigroup} \cite{howe:oscillatorsemigroup} and for $\alpha>1$ 
they have the representation \cite{folland:harmonics:phasespace} 
\begin{equation*}
   \mathcal{L}=e^{-2\pi\text{arcoth}\,\alpha(\Position^2+\Momentum^2)}
\end{equation*}
Thus we have that 
$\mathcal{L}\cdot h_n=e^{-(2n+1)(\text{arcoth}\,\alpha)}h_n$, hence $h_0$ is the optimum
of (\ref{eq:lowerbound}).
The special case $\alpha=1$ can be included by observing that then 
$\BHScat(\tau,\nu)\sim\Amb_{h_0h_0}(\tau,\nu)$. Such pseudo differential operators perform simple
projections, in this case onto the span of $h_0$. Note 
that for $\fourier{\sigma}(\tau,\nu)=\langle \phi,\rho(\tau,\nu)\psi\rangle$ follows
\begin{equation*}
   \begin{split}
      \langle g,\sigma(\Momentum,\Position)\gamma\rangle
      &=\langle\fourier{\sigma},\langle g,\rho(\cdot,\cdot)\gamma\rangle\rangle\\
      &=\langle\langle \phi,\rho(\cdot,\cdot)\psi\rangle,\langle g,\rho(\cdot,\cdot)\gamma\rangle\rangle\\
      &=\langle\Amb_{\phi\psi},\Amb_{g\gamma}\rangle
      =\langle g,\phi\rangle\langle\psi,\gamma\rangle
   \end{split}
\end{equation*}
Thus $\sigma(\Momentum,\Position)$ is a rank one projector 
(orthogonal in the case $\psi=\phi$) if $\langle\phi,\psi\rangle=1$.\\[1em] 
Finally we conclude that for underspread channels the Gaussian pulse shape is an approximate solution of 
(\ref{eq:lowerbound}) which becomes more optimal as the support of $\BHScat$ decreases.
Furthermore the solution is exact for a Gaussian scattering function. 
In the quantum channel context the same arguments hold for coherent states (phase space translated Gaussians).

\appendix
\noindent We will sketch now that the results obtained from the lower bound analysis
will hold with minor restrictions in the direct problem. Writing the CP-map $A(\cdot)$ using 
shift operators gives
\begin{equation*}
   \begin{split}
      A(\Gamma)
      &=\int \Shift_{(\tau-\tau_0,\nu-\nu_0)}\tilde{\Gamma}\Shift^*_{(\tau-\tau_0,\nu-\nu_0)} d\mu
   \end{split}
\end{equation*}
where we introduced again an arbitrary offset $(\tau_0,\nu_0)$ as already done 
in (\ref{eq:lowerbound}) for the lower bound analysis, i.e.
$\tilde{\Gamma}\defeq\Shift_{(\tau_0,\nu_0)}\Gamma\Shift^*_{(\tau_0,\nu_0)}$.
Using again (\ref{eq:shift:localapprox}) gives
\begin{equation*}
   \begin{split}
      A(\Gamma)\approx\int(1+2\pi i\tilde{K}-2\pi^2\tilde{K}^2)
      \tilde{\Gamma}(1-2\pi i\tilde{K}-2\pi^2\tilde{K}^2)d\mu\\      
   \end{split}
\end{equation*}
where the self--adjoint operator 
is now $\tilde{K}\defeq((\nu-\nu_0)\Position-(\tau-\tau_0)\Momentum)$. Hence we get the following 
approximation on the optimization functional
\begin{equation*}
   \begin{split}
      \Trace{\,GA(\Gamma)}&
      \approx\Trace\,\int\Big(\frac{1}{2}(1-4\pi^2 \tilde{K}^2)(\tilde{\Gamma} G+G\tilde{\Gamma})
      +4\pi^2\tilde{K}\tilde{\Gamma} \tilde{K}G\\
      &\hspace*{4em}+2\pi i\tilde{K}[\tilde{\Gamma},G]\Big)d\mu+\order(2)
   \end{split}
\end{equation*}
If we restrict furthermore $g$ and $\gamma$ to be real it can be shown that this will become
\begin{equation*}
   \begin{split}
      \Trace{GA(\Gamma)}\approx
      \Trace\,\int\Big(&(1-4\pi^2 \tilde{K}^2)\tilde{\Gamma}G+4\pi^2\tilde{K}\tilde{\Gamma}\tilde{K}G\\
        &+2\pi i\tilde{K}[\tilde{\Gamma},G]\Big) d\mu
   \end{split}
\end{equation*}
Thus from the latter we can separate the following 
approximated version $A_1(\cdot)$ of the CP-map $A(\cdot)$
\begin{equation*}
   \begin{split}
      A_1(\Gamma)\defeq
      \int\Big(&(1-4\pi^2 \tilde{K}^2)\tilde{\Gamma}+4\pi^2\tilde{K}\tilde{\Gamma} \tilde{K}
      +2\pi i[\tilde{K},\tilde{\Gamma}]\Big)d\mu
   \end{split}
\end{equation*}
In the next steps we will perform the integration of the three integrands. Using
\begin{equation*}
   \begin{split}
      \int \tilde{K}^2\tilde{\Gamma}d\mu=
      &C_{20}\Momentum^2\tilde{\Gamma}+C_{02}\Position^2\tilde{\Gamma}
      -C_{11}(\Momentum\Position\tilde{\Gamma}+\Position\Momentum\tilde{\Gamma})\\
      \int\tilde{K}\tilde{\Gamma} \tilde{K} d\mu=
      &C_{20}\Momentum\tilde{\Gamma}\Momentum+C_{02}\Position\tilde{\Gamma}\Position
      -C_{11}(\Momentum\tilde{\Gamma}\Position+\Position\tilde{\Gamma}\Momentum)\\
      \int[\tilde{K},\tilde{\Gamma}]d\mu=
      &C_{10}[\Momentum,\tilde{\Gamma}]-
      C_{01}[\Position,\tilde{\Gamma}]
   \end{split}
\end{equation*}
we get finally
\begin{equation*}
   \begin{split}
      A_1(\Gamma)
      &=C_{00}
      -4\pi^2\Big(
      C_{20}\Momentum[\Momentum,\tilde{\Gamma}]
      +C_{02}\Position[\Position,\tilde{\Gamma}]\Big)\\
      &+4\pi^2\Big( C_{11}\Momentum[\Position,\tilde{\Gamma}]+
      C_{11}\Position[\Momentum,\tilde{\Gamma}])\Big)\\
      &+2\pi i\Big(C_{10}[\Momentum,\tilde{\Gamma}]-C_{01}[\Position,\tilde{\Gamma}]\Big)
   \end{split}
\end{equation*}
We choose again $(\tau_0,\nu_0)$ such that $C_{10}=C_{01}=0$. Furthermore let us assume 
again for simplicity that the scattering function is separable around $(\tau_0,\nu_0)$ yielding
$C_{11}=0$. Then we will get
\begin{equation*}
   \begin{split}
      A_1(\Gamma)
      &=C_{00}
      -4\pi^2\Big(
      C_{20}\Momentum[\Momentum,\tilde{\Gamma}]
      +C_{02}\Position[\Position,\tilde{\Gamma}]\Big)\\
   \end{split}
\end{equation*}
Next we apply the same dilation procedure used already for the lower bound analysis. 
Let $\tilde{\Gamma}_{\alpha}\defeq d_{\alpha}\tilde{\Gamma}d_{1/\alpha}$.
Using again (\ref{eq:operator:dilation}) gives
\begin{equation*}
   \begin{split}
      A_1(\Gamma)
      &=C_{00}
      -4\pi^2\sqrt{C_{02}C_{20}}\Big(
        \Momentum[\Momentum,\tilde{\Gamma}_\alpha]
        +\Position[\Position,\tilde{\Gamma}_\alpha]\Big)\\
   \end{split}
\end{equation*}
This will result in the same pulse scaling rule as from the lower bound analysis. 
Hence it remaines to show that Gaussians are the right pulses to perform the 
scaling, thus we aim at maximization of $\lVert A_1(\Gamma)\rVert_\infty$ which is 
\begin{equation*}
   \begin{split}
      &\max_{\Gamma} \lVert\kappa/2-\Momentum[\Momentum,\tilde{\Gamma}_\alpha]
      -\Position[\Position,\tilde{\Gamma}_\alpha]\rVert_\infty=\\
      &\max_{\Gamma} \lVert\kappa/2-(\Momentum^2+\Position^2)\tilde{\Gamma}_\alpha + 
      \Momentum\tilde{\Gamma}_\alpha\Momentum+
      \Position\tilde{\Gamma}_\alpha\Position\rVert_\infty\\
   \end{split}
\end{equation*}
where $\kappa=\frac{C_{00}}{2\pi^2\sqrt{C_{02}C_{20}}}$.
It can be shown that for time--frequency symmetric $\tilde{\Gamma}_\alpha$ and $G$ follows 
$\Trace{\,G\Position\tilde{\Gamma}_\alpha\Position}=\Trace{\,G\Momentum\tilde{\Gamma}_\alpha\Momentum}$=0.
With this restriction remains
 \begin{equation*}
   \begin{split}
      &\max_{\Gamma} \lVert\kappa/2-(\Momentum^2+\Position^2)\tilde{\Gamma}_\alpha\rVert_\infty\\
   \end{split}
\end{equation*}
which is maximized by $\tilde{\Gamma}_\alpha$ being a projection onto the eigenspace of $\Momentum^2+\Position^2$
corresponding to the minimal eigenvalue which is again $h_0$. For Gaussian scattering functions in turn 
the calculation in \cite{arxiv:0409063} suggest that this will hold also for $A(\cdot)$.


\bibliographystyle{ieeetr}
\bibliography{references}

\end{document}